\documentstyle[11pt,newpasp,twoside,epsf]{article}
\def\edcomment#1{\iffalse\marginpar{\raggedright\sl#1\/}\else\relax\fi}
\reversemarginpar

\begin{document}
\title{A Keck Adaptive Optics Search for Young Extrasolar Planets}
\author{Denise Kaisler, Ben Zuckerman, and Eric Becklin}
\affil{Department of Physics and Astronomy UCLA, 8371 Math Sciences
  Bldg. Box 951562 Los Angeles, CA 90095-1562, kaisler@astro.ucla.edu}
\author{Bruce Macintosh}
\affil{Institute for Geophysics and Planetary Physics, Lawrence
  Livermore National Labs, 7000 East Ave. L-413, Livermore, CA 94551}

\begin{abstract}
Adaptive optics, large primary mirrors, and careful selection of target stars are         
the keys to ground-based imaging of extrasolar planets. Our near-IR             
survey is capable of identifying exoplanets of 1-10 M${_J}$ within 100 AU of    
young (t$<$60 Myr), nearby (d$<$60 pc) stars. For very young and proximate      
targets such as GJ 803 (12 Myr, 10 pc) we are able to detect exoplanets         
with parameters approaching those of planets our solar system (1 M${_J}$        
at 19 AU, 2 M${_J}$ at 9 AU; 5 M${_J}$ at 5 AU). We have thus far imaged over      
100 stars with  Keck AO. Here we report on our progress, discuss specific       
observing strategies, and present detailed sensitivity limits.
\end{abstract}

\section{Introduction} \label{intsec}

Direct detection is a crucial next step in our understanding of
extrasolar planets. Although the radial velocity technique continues to
generate valuable discoveries (Marcy \& Butler 2000), the composition,  
origins, and essential natures of exoplanets will likely remain hidden 
unless we can study them directly.

To date, we know of over 100 exoplanets, all with semimajor axes $<$ 6
AU. Using the radial velocity technique, detection of exoplanets with
orbital radii similar to those of Saturn, Uranus, and Neptune would require
timelines on the order of decades. However, the existence of these outer planets, 
theories of planetary dynamics (e.g. Rasio \& Ford, 1996), and the
shaping of dusty debris disks (Holland et al 1998; Greaves et al. 1998;
Liou \& Zook 1999, Ozernoy et al. 2000) all lend credence to the idea of
planets with semimajor axes of tens to hundreds of AU.

\section{Strategies for Heightened Sensitivity}

The primary requisites of a workable strategy for detection of extrasolar
planets are high-contrast imaging and subarcsecond resolution. These
capabilities are available via the use of ground-based adaptive optics
(AO) systems or direct imaging from space. A choice of AO in turn
mandates the use of large telescopes, since the time required to achieve
a given S/N ratio varies as the inverse fourth power of mirror diameter
(Olivier et al. 1997). Yet even with Keck AO, planets with semi-major
axes of order tens of AU can be spatially resolved only around stars within
$\sim$ 60 pc of Earth. Current AO sensitivities also rule out the
detection of solar-age planets in reflected light, however, young ($<$ 60 Myr),
warm planets may be seen via their near-infrared emission.

Selection of young targets is effected chiefly through the study of
galactic space motions (U,V,W) in conjunction with excess emissions of
stars. These may appear as X-rays, H$\alpha$ and Ca II H \& K emission,
or far-infrared excesses (e.g. Spangler et al. 2001). Lithium absorption
may also be an important youth indicator. Techniques for identifying
young stars are further described in Zuckerman et al. (2001).  

Once targets have been identified and observed, registration of
long-exposure images and a process called unsharp-masking increase our
sensitivity to point sources. Unsharp-masking is a process in which a
smoothed version of an image is subtracted from the original so as to
improve the contrast between the background noise and small-scale
structure. As can be seen in the next section, unsharp-masking improves
our sensitivity to point sources by as much as two magnitudes within 2''
of a target star.

\section{Project Status} \label{consec}

Our study utilizes the AO system on Keck II (Wizinowich et
al. 2000). Over the past two years, we have observed $\sim$ 100 nearby,
young stars using KCAM, NIRSPEC's slit-view camera (SCAM), and most
recently NIRC2, the first camera optimized for Keck AO. We estimate that
more than half of our target stars are $<$ 100 Myr old. 

Figure 1 shows sensitivity calculations for representative targets from
the SCAM and NIRC2 datasets. As expected, sensitivity improves both for
NIRC2 data and for images which have been unsharp-masked. 

\begin{figure}[!h] 
\plottwo{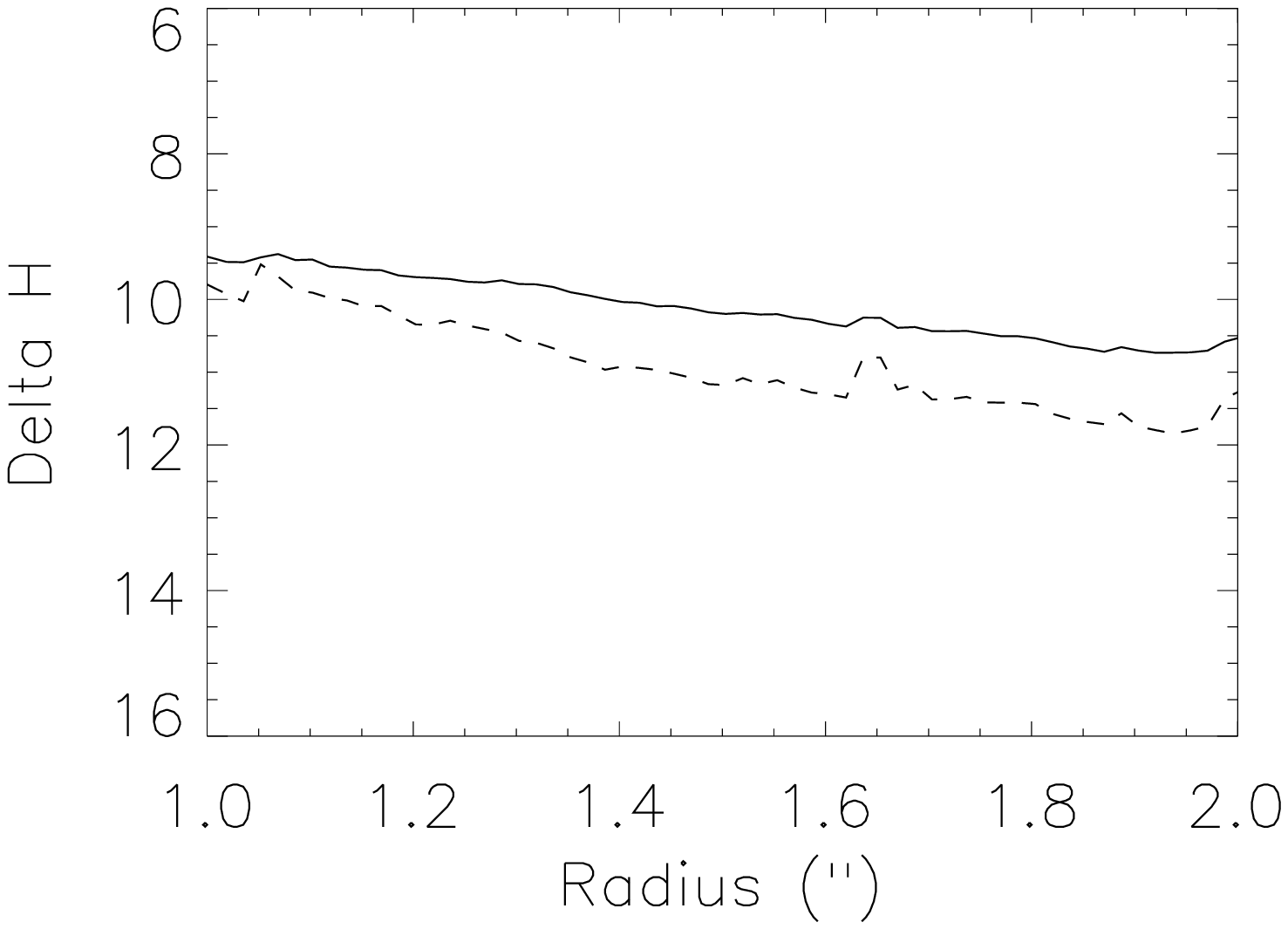}{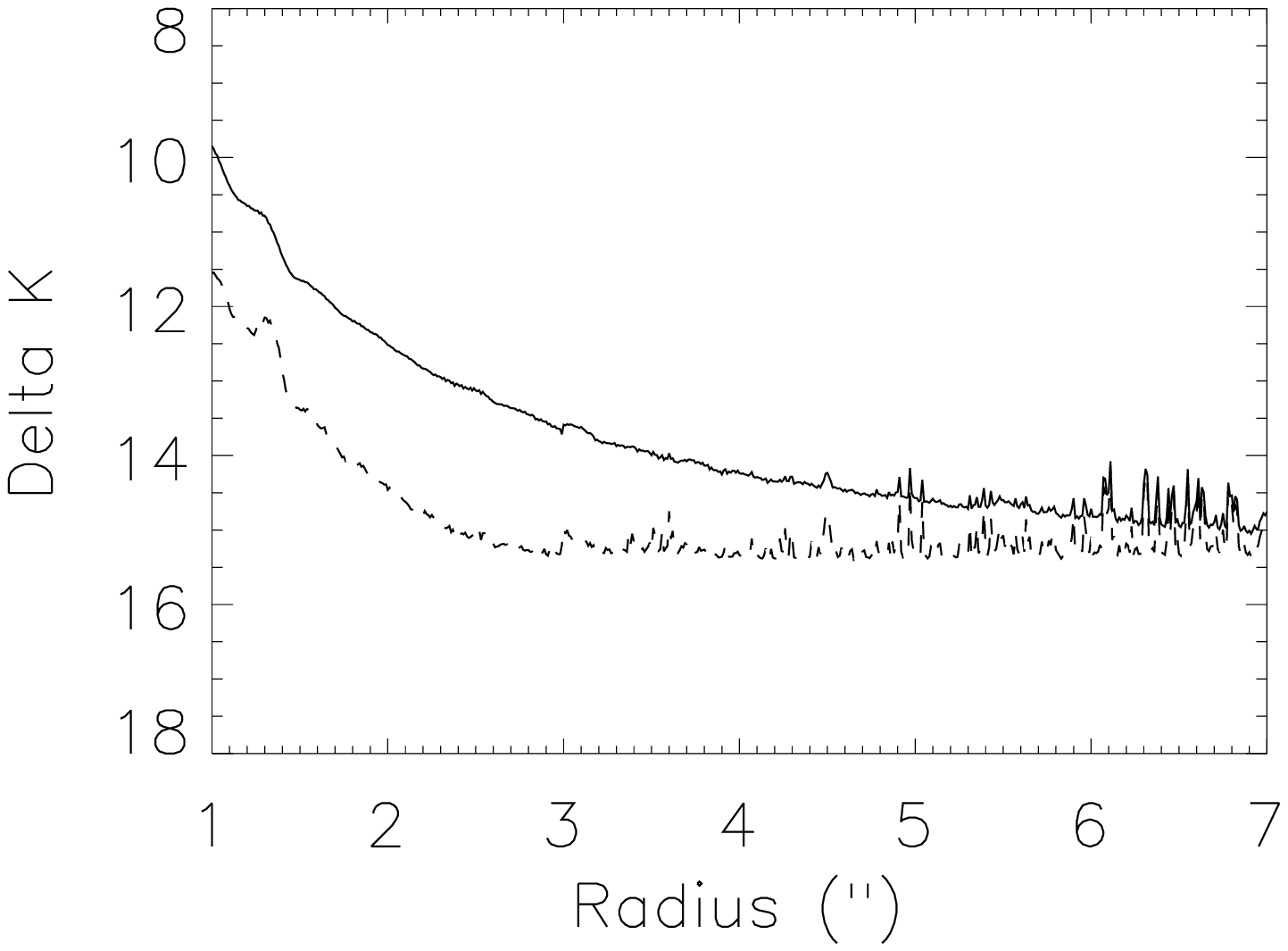}
\caption{Sensitivity profile for (left) GJ 803 (H = 5.2, Strehl = 0.15 at H)
  taken with SCAM and BD+483686 (H = 6.6 , Strehl = 0.45 at K') taken
  with NIRC2. The solid lines represent 5-sigma detection limits for a
  mosaic of sixteen 30-second images. The dashed lines show the
  improvements in sensitivity brought about by unsharp-masking.}
\end{figure}

\begin{table}[!h]
\caption{Planetary Detection Limits at Representative SCAM Stars}
\begin{tabular}{llllllll}
\\
\tableline
\\
Star      &  d  &  age  &  Strehl & Band & 2 M$_J$  & 5 M$_J$ &  10 M${_J}$ \\
          &  (pc)& (Myr)&         && R (AU)  &   R (AU)  &  R (AU)  \\
\\
\tableline
\\
GJ803     &  10 &  12   &  .15    & H &9       &  5   &   3   \\
HD207129  &  16 &  30   &  .10    & H &32      &  16  &    10 \\
HIP 560   &  39 &  12   &  .40    & K'&78      &  35  &    23 \\
HIP82587  &  30 &  30   &  .20    & H &N/D     &  37  &    25 \\
HD159911  &  $\sim$60 &  30   & .40 &  K'    &N/D     &  54  &    30 \\        
\\
\tableline \tableline
\end{tabular}
\end{table}

Using the age of GJ 803, we derived orbital separations at which
this study would detect planets of 2, 5, and 10 Jupiter masses. By
scaling with respect to star age and data quality (measured by
Strehl ratio), we may extend this result to other target stars (Table 1).

Our sensitivity to planets of a given mass varies with such factors as age,
distance, and magnitude of the host star, as well as conditions under
which the data were taken. However, for the best stars in our sample,
of which GJ 803 is an example, this survey is able to detect 1 M${_J}$
planets with orbital radii of only 19 AU.

\acknowledgments{Portions of this work were performed under the auspices
  of the U.S. Department of Energy, National Nuclear Security
  Administration by the University of California, Lawrence Livermore
  National Laboratory under contract No. W-7405-Eng-48. This work was
  also supported by the National Science Foundation Science and
  Technology Center for Adaptive Optics, managed by the University of
  California at Santa Cruz under cooperative agreement No.  AST -
  9876783. Additional funding was provided by a NASA grant to UCLA.}


\end{document}